\documentclass[nopreprintline, authoryear,12pt]{elsarticle}
\usepackage{epsfig}

\usepackage{amssymb}

\usepackage[%ps2pdf,%
a4paper=true,%
breaklinks=true,%
colorlinks=true,%
pdfauthor={T. J. Brandt et al.},%
pdftitle={A View of Supernova Remnant CTB 37A with the Fermi Gamma-ray Space Telescope}%
]{hyperref}

\usepackage{subfig}

\journal{Advances in Space Research}

\newcommand{\apjl}{ApJL }
\newcommand{\apj}{ApJ }
\newcommand{\aj}{AJ }

\newcommand{\apjs}{ApJS }

\newcommand{\aap}{AAP }

\begin{document}

%%%%%%%%%%%%%%%%%%%%%%%%%%%%%%%%%%%%%%%%%%%%%%%%%%%%%%%%%%%%%%%%%%%%%%%%%%%%%
%% Frontmatter
\begin{frontmatter}


% Use the tnoteref command within \title and fnref within \author or \address for footnotes;
% use the corref command within \author for corresponding author footnotes;
% \author{Name\corref{cor1}\fnref{label2}}
% \address{Address\fnref{label3}}

\title{A View of Supernova Remnant CTB 37A with the Fermi Gamma-ray Space Telescope} %\tnoteref{footnote1}}
%\tnotetext[footnote1]{This template can be used for all publications in Advances in Space Research.}

% Use optional labels to link authors explicitly to addresses:

\author{T. J. Brandt\corref{cor}\fnref{fn:corrAuthor}}
\address{9, av du Colonel Roche BP 44346 / 31028 Toulouse Cedex 4}
\cortext[cor]{Corresponding author}
\fntext[fn:corrAuthor]{Universit\'e de Toulouse, UPS-OMP, IRAP and CNRS, IRAP}
\ead{brandt@cesr.fr}

% Url can be given like this:
% \ead[url]{http://www.elsevier.com/wps/find/authorsview.authors/latex}

%\author{J. Knodlseder, P. Jean \fnref{footnote3}}
%\address{Address of the second and third authors}
%\fntext[footnote3]{Additional information about the second and third authors}
%\ead{more@email.addresses}

\author{on behalf of the Fermi Large Area Telescope Collaboration}%\fnref{footnote4}}
%\address{Address of the co-authors}
%\fntext[footnote4]{Additional information about the co-authors}
%\ead{more@email.addresses}

\begin{abstract}

Supernovae and their remnants have long been favored as cosmic ray accelerators. Recent data from the Fermi Gamma-ray Space Telescope has given us an improved window into such sources, including the remnant CTB 37A. Using the Fermi Large Area Telescope, we found significant %, extended Removed due to Josh's sourcelike finding of same extension but significance of ~3sigma for P7 with ~2 years data
gamma-ray emission coincident with the remnant, which also emits in radio, X-ray, and very high energy gamma-rays. We modeled the multiwavelength spectrum using a combination of hadronic and leptonic emission with reasonable parameter values and determined that CTB 37A is a potential cosmic ray accelerator commensurate with direct observations. By assembling statistically significant populations of such objects, we will be able to more fully illuminate the mystery of cosmic ray origins. 

%present a combination hadronic and leptonic model with reasonable parameter values of the multiwavelength spectrum using  
%We present the results of an analysis of Fermi data of one such source CTB 37A and discuss potential implications for cosmic ray acceleration.

%\fnref{fn:test}
%\fntext[fn:test]{Additional information regarding the corresponding author}

%This is a template for manuscripts for the journal Advances in Space Research (ASR). It shows title, footnotes, abstract, keywords, text, one figure, one equation, one table, citations and references.
%We hope authors of ASR will find this template useful.
%For detailed instructions for use of the documentstyle {\it elsarticle} and several features of preparing articles with the {\it elsarticle} documentstyle please see \href{http://www.elsevier.com/latex}{LaTeX section in the Elsevier AUTHORS HOME}.
%You can also \href{http://www.elsevier.com/framework_authors/misc/elsarticle.zip}{download the class file} from this website.

\end{abstract}


\end{frontmatter}

\parindent=0.5 cm

%%%%%%%%%%%%%%%%%%%%%%%%%%%%%%%%%%%%%%%%%%%%%%%%%%%%%%%%%%%%%%%%%%%%%%%%%%%%%
%% Main text
\section{Introduction}

% You can simply replace the text/figures/tables in this file
% with your own material and process it to produce a referee version
% or camera ready version of your manuscript.

The mystery of the origin of cosmic rays (CRs) has lasted for nearly 100 years. Energetics arguments strongly suggest supernovae and their remnants (SNe \& SNRs) may be a source of Galactic CRs. By observing such potential CR accelerators in multiple wavelengths, we can constrain their particle populations, the acceleration processes they undergo, and thus the sources' ability to accelerate CRs up to some of the highest energies observed. 

We do so for the SNR CTB 37A, which we detect with the Fermi Gamma-ray Space Telescope (Fermi). The slightly extended emission coincides with the nominal position of CTB 37A, existing radio and X-ray data, and H.E.S.S. very high energy gamma-ray observations. Previous studies suggest that the SNR may be interacting with molecular clouds seen in CO to be positionally coincident with the remnant \citep{ReynosoMagnumCO}. The clouds are also kinematically associated with several OH $1720\,$MHz masers, easily explained through collisional shock excitation \citep{Frail96Masers}. In this way, SNR-cloud interactions may be another source of both CRs and gamma-rays emitted by CTB 37A. Motivated by these observations, we present a model of the multiwavelength spectrum which both reasonably reproduces the Fermi data and suggests that this SNR could be accelerating CRs up to the observed energies.
%An independent analysis of CTB 37A (Castro and Slane, 2010) reaches similar conclusions.
An independent analysis of CTB 37A \citep{CastroSlane2010} reaches similar conclusions.
%An independent analysis of CTB 37A by Castro and Slane (2011) reaches similar conclusions.

% Combining these observations, we present a model motivated by the multiwavelength spectrum 

\section{Fermi LAT Analysis: Methods and Results}\label{sec:AnalysisMethodResults}

The Fermi Gamma-ray Space Telescope, launched 11 June 2008, contains the Gamma-ray Burst Monitor and the Large Area Telescope (LAT). The LAT's silicon tracker modules and Cesium Iodide calorimeter are sensitive to photons in a broad energy range ($\sim 0.02$ to $>300$\,GeV) with a large effective area ($\sim 8,000\,\mathrm{cm}^2$ for $E>1$\,GeV for on-axis events) over a large field of view ($\sim 2.4$\,sr). The front (first $12$) tracker planes have thin Tungsten converter foils enabling a smaller point spread function (PSF) while the back (last $4$) planes' thicker converters permit a larger effective area.

We used approximately $18$\,months of the standard, low background {\it diffuse} class LAT data, collected from 8 Aug 2008 to 12 Feb 2010. A zenith cut of $105^{\circ}$ minimizes Earth albedo gamma-rays while limiting the energy range to $0.2 < E < 50$\,GeV minimizes the systematic error. We used the standard science tools (v9r15p6) and instrument response functions (IRFs) (P6\_V3)\footnote{Available at the Fermi Science Support Center. \url{http://fermi.gsfc.nasa.gov/ssc}} %\fnref{fn:FSSC} 
to fit a $4.5^{\circ}$ region of interest (ROI) around the known location of the source \citep{GreensCat}. In particular, we used the {\it gtlike} binned maximum likelihood fit to optimize the spectral parameters of all sources from the 1FGL catalog \citep{1FGLCat} in the ROI, iterating to include only significant sources ($>4\,\sigma$). As CTB 37A is quite near the Galactic plane, the smaller ROI limited the number of sources included in the {\it gtlike} fit of the region to a more manageable number while still containing a sufficient number of PSF length scales. We also included the standard isotropic\footnote{All contributions whose spatial distribution is assumed to be isotropic: the extragalactic background, unresolved sources, and the instrumental background.} %\fnref{fn:isoBkg}
and Galactic diffuse models: isotropic\_iem\_v02 and gll\_iem\_v02, respectively\fnref{fn:FSSC}. % \cite{}cite??. 
Further details of the LAT instrument and data reduction may be found in \cite{AtwoodLAT}. %\footnote{this is a test of fnoting}
\fntext[fn:FSSC]{Available at the Fermi Science Support Center: \url{http://fermi.gsfc.nasa.gov/ssc}}
\fntext[fn:isoBkg]{All contributions whose spatial distribution is assumed to be isotropic: the extragalactic background, unresolved sources, and the instrumental background.}

\subsection{Detection}\label{sec:Detection}

\begin{figure}
\begin{center}
\includegraphics*[width=10cm]{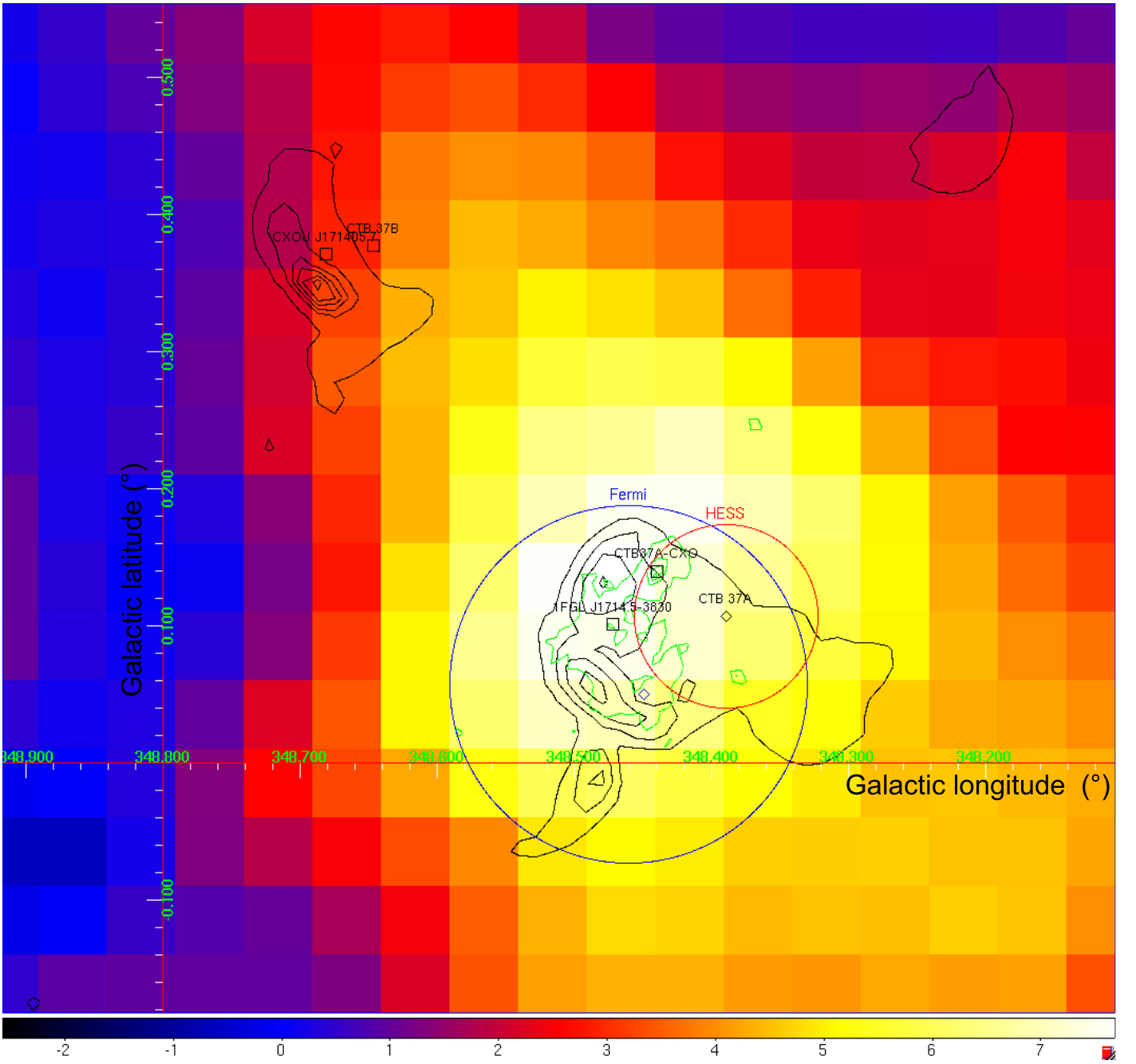}
\end{center}
\caption{The Fermi-LAT residual map (color scale) in the energy range $0.2 - 50\,$GeV shows an excess clearly associated with both the nominal position (black diamond \citep{GreensCat}) and radio emission (southern black contours \citep{Kassim91}) of CTB 37A and not CTB 37B (northern black contours). The Fermi-LAT emission's most likely position and Gaussian extension (blue circle) also coincides with the H.E.S.S. source (red circle \citep{HESS_CTBA}) as well as the XMM-Newton MOS1 detector X-ray data (green contours) in the energy range $0.2-10\,$keV. We performed a blind search for pulsations at the four positions indicated with black squares: at the 1FGL catalog position and the nearby possible X-ray source for CTB 37A and at the nominal and coincident X-ray source positions for CTB 37B.} \label{fig:resMap}
\end{figure}

This analysis yielded an $18.6\,\sigma$ detection of a point source coincident with the nominal CTB 37A position, as seen in Figure \ref{fig:resMap}, a map of residual Fermi-LAT emission after subtracting all the nearby, fit sources and diffuse backgrounds.  For this detection, we used {\it gtlike} to quantify the spectral shape with a power law. Discussion of the fit with another reasonable spectral form, the exponentially cutoff power law, may be found in \ref{sec:Var}.  In Section \ref{sec:MWSpec} we explore a physically motivated model. %fit the spectral parameters of

\subsection{Diffuse Model}\label{sec:DiffModel}

Since CTB 37A lies just above the Galactic plane we ensured that the particular Galactic diffuse background employed was not biasing our detection by examining the global likelihood of three different diffuse models including several variations for the two most reasonable models. As expected, the DIRBE $60\,\mu$m infrared map, mainly tracing thermal emission from e.g., dust heated by stars, including interplanetary dust heated by the Sun, poorly represents the diffuse Galactic gamma-ray background and so led to a low global fit likelihood relative to the standard model. %(see Section \ref{sec:AnalysisMethod}). 
Using a variation on the standard model, fitting it just to a region around the Galactic center ($(l,b) = \pm (20^{\circ}, 20^{\circ})$), also did not substantially improve the global fit to the ROI. Neither including a faint pulsar in the region found after the 1FGL catalog was published nor modulating the original Galactic diffuse model by a power law spectral shape significantly improved the global fit, particularly considering the addition of extra parameters in each case. 

The global fit did improve when using a variation of the Galactic diffuse model with finer spatial resolution based on $19$ months' data. %by at least $12\,\sigma$ for all variations. %, as expected for an improved diffuse model. 
The fit for this model further improved as we included sources from the 1FGL catalog which had a lower significance ($>3\,\sigma$) %\fnref{fn:TS10} 
in the initial fit and as we modulated the Galactic diffuse model by a power law in energy. Combining these two variations and further including in the combination the faint pulsar both yielded moderate improvements in the global likelihood over any one variation singly. %(of order $4\sigma$) 
While the $19$ months' model variations yielded the best global fit, they remain under development, and the other models are either worse or show negligible improvement in the global fit relative to the standard diffuse model. Further, none of the reasonable models significantly influence the results for the point source at the nominal CTB 37A position. The minimum source significance was within $8\%$ of the average of all reasonable variations and the best-fit power law index remained within $3\%$ of the average. We therefore conclude that the standard Galactic diffuse model used in the analysis is adequate to the task and further that neither it nor its variations greatly affect our results.

%The $19$ months' model yielded the minimum source significance of $\sim 16.6\,\sigma$ for a point source at the nominal CTB 37A position, only marginally less than that for the standard diffuse model. Further, the point source's best-fit power law index was within $3\%$ of the average for all reaonable diffuse model variations.
%Since the $19$ months' diffuse model variation is still under development, the other models are either worse or show negligible improvement in the global fit relative to the standard diffuse model, and the source itself is quite significant under all reasonable variations of the diffuse model, we conclude that the standard Galactic diffuse model used in the analysis is adequate to the task and further that neither it nor its variations greatly affect our results.
%\fntext[fn:TS10]{$TS>10$ rather than $>20$ (**prob delete)}

\subsection{Location and Extension}\label{sec:LocExtMethod}

Having determined the existence of a significant source in the vicinity of the CTB 37 complex, we simultaneously fit for the most likely position and extension of the source. To do so, we stepped around a spatial grid to determine the most likely position for a given (Gaussian) extension, recentering the grid as needed to more completely included the largest extensions ($0.3^{\circ} - 0.5^{\circ}$). We interpolated the most probable extension bounded by a point source and $0.5^{\circ}$ and then iterated over a finer range of extensions on a finer grid spacing (from $1.0^{\circ}$ to $0.1^{\circ}$ on a $5\times 5$ grid) to improve the final location, extension, and their respective statistical errors. The final location is that associated with the most likely extension. %Figure \ref{fig:extGrid} shows the most probable extension, seen as the peak of the parabolic fit to extension versus TS [Include this plot?? probably not.....].
This method was previously employed in e.g. \citet{JurgenLMC}. 
We also verified the results using the alternate tool {\it Sourcelike} \citep[e.g.]{IC443LAT}.% (better reference??). 
We obtained an identical extension as for the grid method, and further determined that a uniform disk was no more likely than a Gaussian extension. 

We thus found a position for the Fermi-LAT source of RA $= 258.68^{\circ} \pm 0.05^{\circ}_{stat} \pm 0.006^{\circ}_{sys}$, DEC $= -38.54^{\circ} \pm 0.04^{\circ}_{stat} \pm 0.02^{\circ}_{sys}$. The symmetrical Gaussian with an extension of $\sigma = 0.13^{\circ} \pm 0.02^{\circ}_{stat} \pm 0.06^{\circ}_{sys}$ was most likely, with a significance of $\sim 4.5\,\sigma$. The position and extension can be seen in Figure \ref{fig:resMap} overlaid on the Fermi-LAT residual map, along with the radio contours for CTB 37A and B and the position and extension of the H.E.S.S. source coincident with CTB 37A. We clearly see that the Fermi-LAT source is coincident with both the CTB 37A radio contours and the H.E.S.S. source, and further that it is inconsistent with CTB 37B. 

We further checked that the Galactic diffuse model and possible nearby sources did not affect our extension and location determination by performing the same analysis with three of the most reasonable models' variations: including the faint pulsar in the standard model, including less-significant sources from the 1FGL catalog in the $19$ months' diffuse model variation, and including the less-significant sources and modulating the $19$ months variation by a power law in energy. In all three cases, the position and extension found were within statistical errors of the originally determined position and extension. Thus we conservatively take the extreme difference as the systematic error due to the diffuse model, which is only $\sim 30\%$ of the extension and well below the statistical error ($< \pm 0.015^{\circ}$) for the position. %These are well within the systematic error from the PSF.% (right??).  -->
 The diffuse model systematic error thus derived is smaller than the PSF.

We also examined the extension and localization using subsets of events with a nominally better PSF: those at higher energy ($2-50$\,GeV) and those photons which convert in the front of the tracker, where the foils are thinnest and thus cause the least scatter. In both cases, the position remained stable and the extension within the statistical errors. 
%We used the error on the PSF, derived from a detector simulation verified in accelerator tests \citep{AtwoodLAT}, to place an upper limit on the 
We again use the maximum difference to estimate the order of magnitude of systematic error from the PSF, which we combined in quadrature with the systematic error from the diffuse model.

\subsection{Variability and Pulsations}\label{sec:Var}
Fermi GST has identified emission from over $70$ pulsars and variable emission from over $180$ blazars. As pulsing or variable emission can strongly influence the results found for a steady source, we performed two searches of the Fermi-LAT source coincident with the CTB 37 complex: a search for long-term variability in the light curve and a blind pulsation search of the most likely candidates in the region. We found no significant variability in the 2-week binned light curves when fitting the spectrum with either a power law or a power law with exponential cutoff (PLEC) spectral shape. All fluxes remained within error of $1\,\sigma$ of the average except for the two lowest, whose errors are systematically underestimated due to low statistics, and the third highest flux, which had an unusually small error. Replacing these with the errors determined for their nearest neighbors (in flux) yields a reduced $\chi^2$ compared to a constant, average flux of $\chi^2/d.o.f. \approx 1.4$ and $5.7$ for the power law and PLEC, respectively. Thus, we do not observe long-term variability. 

To check for pulsar-like emission, we performed a blind pulsation search at four likely locations coincident with the CTB 37 complex, seen as squares in Figure \ref{fig:resMap}, using 18 months diffuse events with $E \geq 300$\,MeV in a ROI of $r \leq 0.8^{\circ}$. We employed a differencing method with a maximum frequency of $64$\,Hz, a maximum coherency window of $524,288$\,seconds, and a spin down frequency ($\dot{F}$) ranging from $0$ to $3.86 \times 10^{-10}$, equal to that of the Crab.  %\number{524,288}
The 1FGL catalog source J1714.5-3830, coincident with the Fermi-LAT source and CTB 37A radio contours, and the X-ray source CXOJ171419.8m383023, coincident with the Fermi-LAT, radio, and H.E.S.S. CTB 37A sources, showed no pulsations. We also examined the nominal position of CTB 37B (RA $= 258.49$, DEC $= -38.20$) and the nearby X-ray source CXOU\_J171405.7, more coincident with CTB 37B's radio contours, for pulsations, likewise finding none. Further details on the Fermi-LAT blind pulsation search may be found in \cite{BlindPulseSearch}.

%We checked for pulsations in four likely locations coincident with the CTB 37 complex, seen as squares in Figure \ref{fig:resMap}. The 1FGL catalog source J1714.5-3830, coincident with the Fermi source and CTB 37A radio contours, and the X-ray source CXOJ171419.8m383023, coincident with the Fermi, radio, and H.E.S.S. CTB 37A sources, showed no pulsations. We also examined the nominal position of CTB 37B (RA $= 258.49167$, DEC $= -38.20000$) and the nearby X-ray source CXOU\_J171405.7, more coincident with CTB 37B's radio contours, for pulsations, finding none. 

We also find the Fermi-LAT source coincident with CTB 37A inconsistent with a pulsar hypothesis as the best-fit spectral shape, determined with {\it gtlike} to be a PLEC, while consistent with the standard pulsar one, shows only a $\sim 1.5\sigma$ improvement over the simple power law while adding an extra degree of freedom. Further, the best fit cutoff energy $\mathrm{E}_{\mathrm{cut}} = 28.2_{-10.5-21.6}^{+29.8+67.8}\,$GeV is both an order of magnitude higher than typical pulsar cutoff energies (e.g. \cite{PulsarCatalog}) and moreover is also poorly fit, as evidenced by the large errors which are consistent with no cutoff detection. Thus, we find it unlikely that the detected emission comes from a pulsar.

\subsection{Spectrum}\label{sec:FermiSpec}

\begin{figure}
\begin{center}
\includegraphics*[width=12cm]{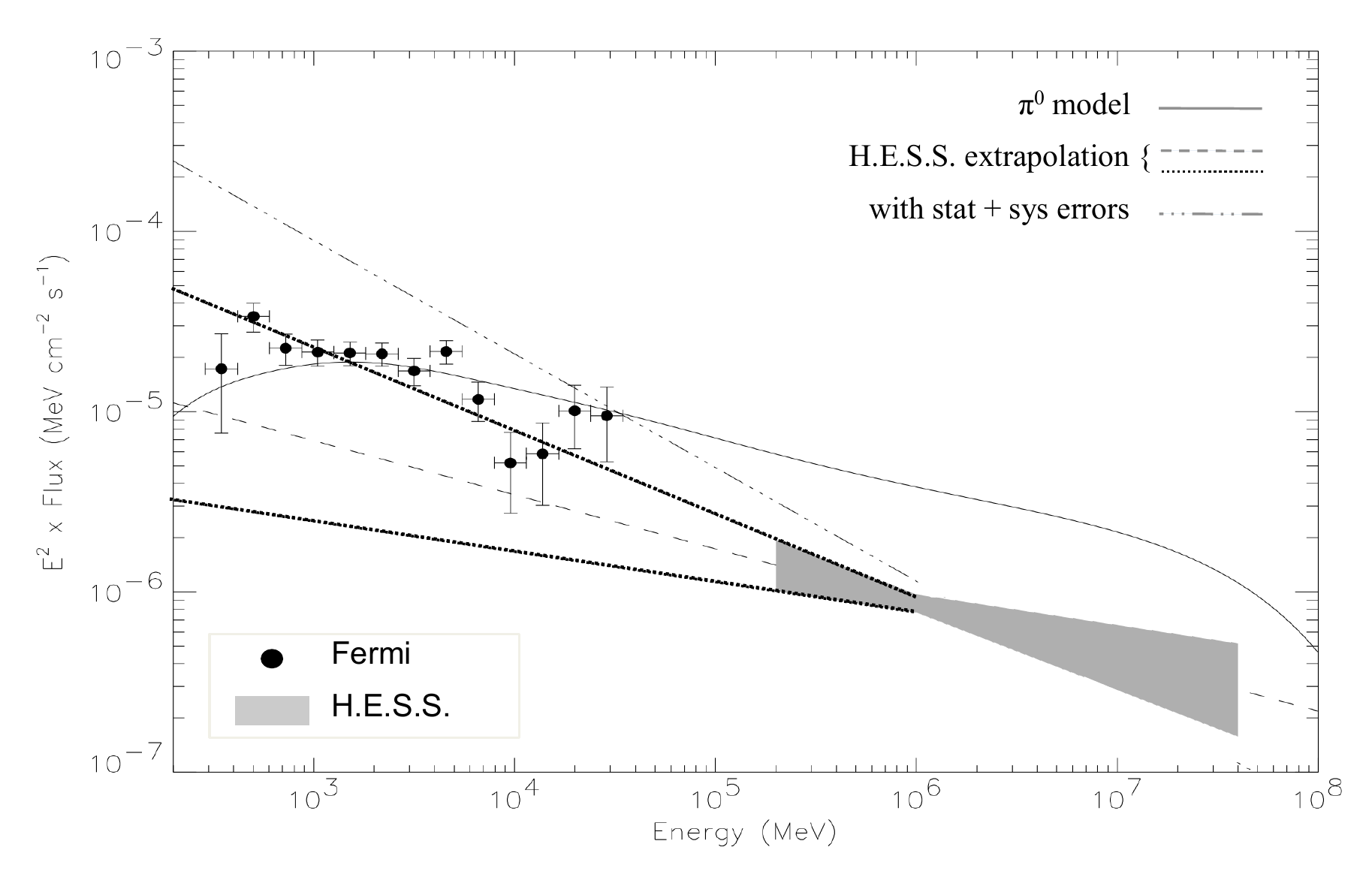}
\end{center}
\caption{The Fermi-LAT (points, $1 \sigma$ statistical error bars) and H.E.S.S. (grey shaded bow tie \citep{HESS_CTBA}) spectra from emission coincident with SNR CTB 37A. The dashed line, an extrapolation of the H.E.S.S. data into the Fermi-LAT energy range, clearly underpredicts the Fermi-LAT data. The discrepancy is somewhat ameliorated by including H.E.S.S. statistical errors (dotted lines) and is eliminated by very conservatively summing the H.E.S.S. statistical and systematic errors (dot-dashed line). 
%Including statistical errors (dotted lines) somewhat ameliorates the discrepancy, while very conservatively summing the statistical and systematic errors (dot-dashed line) eliminates the difference. 
The spectral tension remains after instead normalizing the best-fit H.E.S.S. hadron model to the Fermi-LAT data (solid curve). }\label{fig:spec}
\end{figure}

To create the Fermi-LAT spectral points shown in Figure \ref{fig:spec}, we used {\it gtlike} to fit power laws to the data within each of $15$ logarithmically-spaced energy bins. The narrowness of the bins in energy space ensures that the exact functional form and parameters of the power law do not substantially influence the values of the points themselves. %[< systematic error]. 
We do not show the lowest ($0.2-0.3$\,GeV) and highest ($30-50$\,GeV) energy bins due to insufficient statistics. %Upper limits are not yet significantly constraining.
The variations in the Fermi-LAT data points are not overwhelmingly significant in light of systematic errors for this standard Fermi-LAT analysis. 

\section{Multiwavelength Spectrum}\label{sec:MWSpec}
By observing objects in multiple wavelengths, we are able to gain a more complete understanding of their internal workings, and SNRs are no exception. 
Figure \ref{fig:spec} shows the gamma-ray spectrum for the Fermi-LAT source associated with CTB 37A. 

\subsection{High Energy Gamma-rays}\label{sec:gammaModel}

The gamma-ray band is particularly sensitive to possible CR acceleration through the $\pi^0$ decay channel, initiated by the interaction of high energy hadrons, typically taken to be protons. 
In 2008, H.E.S.S. detected very high energy gamma-rays coincident with CTB 37A. That data, in combination with radio and X-ray analysis, suggested that a hadron-dominated emission scenario was more likely than a leptonic one, though the latter could not be excluded \citep{HESS_CTBA}. %, corresponding to CR acceleration,

As the Fermi-LAT and H.E.S.S. energy ranges are complementary and the sources themselves coincident, we examined the extrapolation of the H.E.S.S. data into the Fermi-LAT range. The dashed line in Figure \ref{fig:spec}, showing this extrapolation, clearly underpredicts the Fermi-LAT measured spectrum. Including statistical errors somewhat ameliorates the discrepancy, though the global fit likelihood remains more than $8\sigma$ worse than freely fitting the source. Very conservatively directly summing the statistical and systematic errors eliminates the difference, as seen from the dotted and dot-dashed lines, respectively.

If instead we use {\it gtlike} to fit the Fermi-LAT data with the H.E.S.S.-indicated hadronic emission model (from $\pi^0$ decay) as seen in Figure \ref{fig:spec} (solid curve), the Fermi-LAT data suggest that H.E.S.S. should observe more emission than it does. To generate the model, we used H.E.S.S. parameters for the SNR at $11.3\,$kpc interacting with clouds of gas mass $\mathrm{N}_{\mathrm{H}} = 6.7 \times 10^4\,\mathrm{M}_{\odot}$ with a power law index of $\gamma = 2.30$ \citep{HESS_CTBA}. While the discrepancy may arise from statistical or systematic errors or differences in source extension as the H.E.S.S. source is slightly smaller than and marginally offset from the Fermi-LAT source, it may also indicate two different spectral components. To determine if the latter is an actual possibility, we extended our model to include the radio and X-ray data.

\subsection{A Hadronic + Leptonic Model}\label{sec:MWModel}

\begin{figure}
%\begin{center}
%\includegraphics*[width=12cm]{MWSpec_PrelimModel.pdf}
%\end{center}
\centering
  \subfloat[]{\label{fig:MWSpecFull}\includegraphics[width=10cm, trim=.25cm .4cm .82cm .55cm, clip]{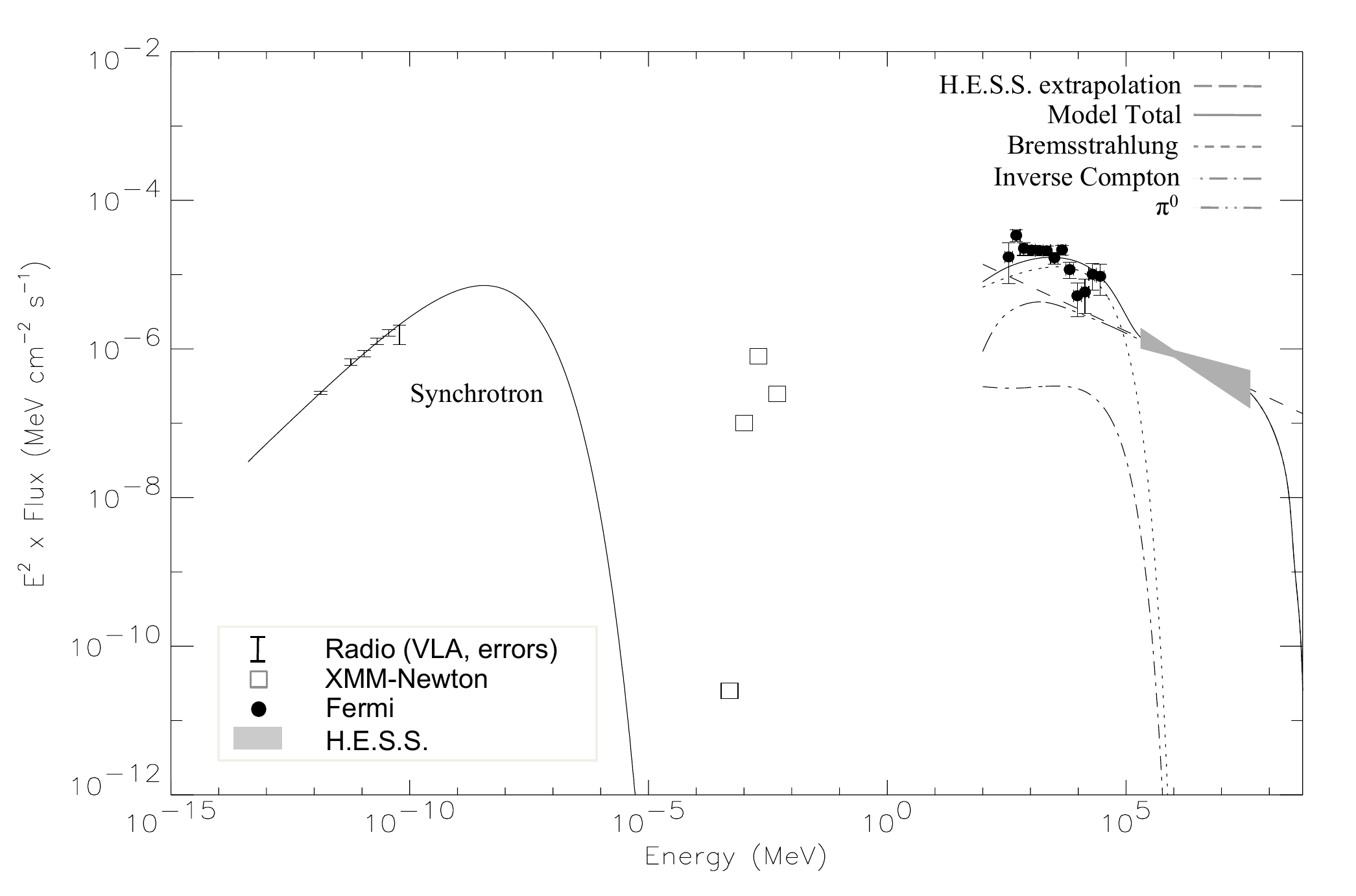}}       \\         
  \subfloat[]{\label{fig:MWSpecZoom}\includegraphics[width=10cm, trim=.25cm .4cm .82cm .55cm, clip]{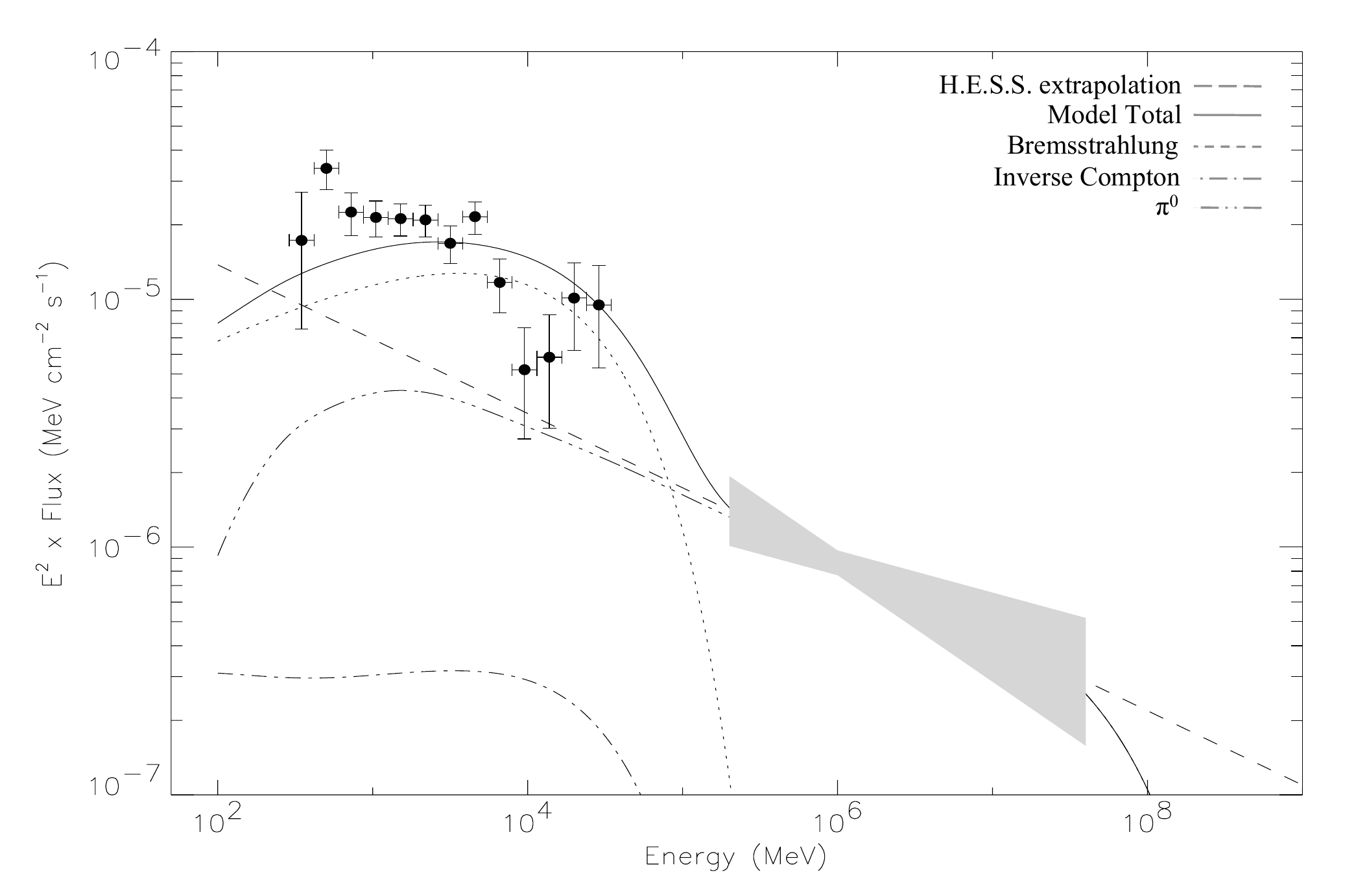}}   %_HEzoom
\caption{\ref{fig:MWSpecFull}. The multiwavelength spectrum of emission associated with CTB 37A, spanning from radio to very high energy gamma-ray emission. We fit the radio data \citep{Kassim91} with a synchrotron model (solid line) having a $20\mu$G magnetic field, determining the lepton population's spectral index and normalization, having assumed a power law with exponential cutoff energy of $50\,$GeV. 
\ref{fig:MWSpecZoom}. We then model the lepton population's bremsstrahlung (dashed line) and inverse Compton (dot-dashed line) emission. We adjusted the hadron population's pion emission (triple-dot-dashed line) from proton-proton interaction in the same ambient clouds as used for the leptons to produce bremsstrahlung to approximately fit the H.E.S.S. data (grey bow tie) \citep{HESS_CTBA}. The total modeled emission (solid line at high energies) relatively well reproduces the Fermi-LAT data without further adjustments. The XMM-Newton X-ray data from the MOS1 and 2 and PN instruments (open squares) constrains the maximum synchrotron emission at the highest energies.} 
\label{fig:MWSpec}
\end{figure}

We fit a one-zone model using reasonable values and containing both hadronic ($\pi^0$ decay) and leptonic (synchrotron, bremsstrahlung, and inverse Compton) emission to the multiwavelength data coincident with SNR CTB 37A, as shown in Figure \ref{fig:MWSpec}. 

We selected radio data tabulated by \cite{Kassim91} with frequencies above $200$\,MHz, where absorption has a minimal effect, and fluxes both with errors and consistent with other measurements. For the $8800.0$\,MHz band, we fixed the error at $10\%$ of the total value, which is the same order of magnitude and slighlty larger than the errors in the other bands.% and does*** not significantly alter the final fit. 

In a reasonable $20\mu$G magnetic field\footnote{Measurements of Zeeman splitting in OH masers coincident with CTB 37A provide an upper limit on the magnetic field magnitude of $1.5\,$mG \citep{BroganZeeman}. Such OH masers typically occur in molecular clouds carrying ambient magnetic fields which have been compressed by the SNR shock passage \citep{BroganZeeman}. Thus, to obtain the value used, we scaled the field by the size of the cloud.}, %\fnref{fn:Bfield}, 
leptons produce radio-band synchrotron emission. We calculated this leptonic model component according to the method of \citet{Ginzburg} and \citet{GhiselliniGuilbertSvensson} and assuming an electron population following a standard exponentially cutoff power law with a typical cutoff energy of $\mathrm{E}_{\mathrm{cut}} = 50\,$GeV. 
We fit the synchrotron component to the radio data, providing the lepton population spectral index and normalization of $\gamma_{\mathrm{e}} = -1.75$ and $\mathrm{N}_{\mathrm{e}} = 3.66\,$e/s/$\mathrm{cm}^2$/GeV/sr ($\sim 219$ times that of the local CR electron spectrum) at $1\,$GeV, respectively. 
\fntext[fn:Bfield]{Measurements of Zeeman splitting in OH masers, where a SNR shock compressed a molecular cloud carrying ambient magnetic fields, coincident with CTB 37A provide an upper limit on the magnetic field magnitude of $1.5\,$mG \citep{BroganZeeman}. To obtain the value used, we scaled the field by the size of the cloud.}

Leptons in this population may also interact with molecular clouds, producing bremsstrahlung emission proportional to the gas mass of the clouds in the region, estimated at $\mathrm{M}_{\mathrm{H}} = 6.5 \times 10^4\,\mathrm{M}_{\odot}$ for the northern and central clouds of \citet{ReynosoMagnumCO}, which are those most consistent with the Fermi-LAT emission. Leptons may also inverse Compton scatter off local starlight and the cosmic microwave background (e.g. as measured by WMAP), producing high energy photons. We calculated the inverse Compton emission using the method described by \citet{BlumenthalGould} with optical and infrared interstellar radiation fields from \citet{Porter2008}.

%Pierre: The inverse Compton component was calculated using the method described by Blumenthal & Gould (1970) using the optical and infrared interstellar radiation fields from Porter et al. (2008).
%We calculate the pion production by proton-proton interactions following the prescription of Kamae et al. (2006). 

We combined the leptonic emission components with a hadronic model in which a high energy proton population, following a power law with index of $-2.30$, interacts with ambient protons, producing $\pi^0$s which decay to two photons, calculated following the prescription of \citet{Kamae2006}. Using the same gas mass as for the bremsstrahlung component, we renormalized the hadronic gamma-ray flux to approximately fit the H.E.S.S. data, determining a CR proton enhancement factor of $\sim 11.6$ with respect to the local CR proton spectrum. 
The protons in this steady state model have a total energy of $\sim 1.2 \times 10^{50}\,$ergs, implying a very reasonable $\sim 12\%$ conversion efficiency for a canonical SN kinetic energy of $10^{51}\,$ergs and in good agreement with the H.E.S.S. prediction \citep{HESS_CTBA}. The total lepton energy, at $\sim 5 \times 10^{49}\,$ergs, is a reasonable roughly half the hadron energy. The proton energy corresponding to the maximum H.E.S.S. energy, $\mathrm{E}_{\mathrm{p,max}}$ between $10^{14}\,$eV and $10^{15}\,$eV \citep{HESS_CTBA}, is commensurate with the maximum energy expected from SNe for particles of charge $Z=1$ \citep{myThesis} and with the \textquotedblleft knee\textquotedblright or change in CR spectral slope at $10^{15}\,$eV.

%We approximately fit the H.E.S.S. emission by varying the gas mass by less than a factor of $10$ (hence brem $\mathrm{N}_{\mathrm{H}} = 6.5$ and not $6.7$ as in HESS paper?) ??Pierre, is this right?? is it the same mass as used for the brem emission?? what is the final mass? why wasn't it just by varying the proton density (normalization) and not introducing another d.o.f.? 

Figure \ref{fig:MWSpec} shows the result of fitting this model to the multiwavelength data. The lepton population reproduces the radio-measured synchrotron emission while the combination of $\pi^0$ decay and bremsstrahlung emission reproduce the Fermi-LAT data, notably not previously fit, relatively well with no further adjustments. The inverse Compton component contributes at only about the $1\%$ level. From this we anticipate obtaining similar values when performing a complete fit of all the data, allowing the magnetic field strength as well as the proton and electron populations' parameters ($\mathrm{N}_{\mathrm{p}}$, $\gamma_{\mathrm{p}}$, $\mathrm{N}_{\mathrm{e}}$, $\gamma_{\mathrm{e}}$, $\mathrm{E}_{\mathrm{e,cut}}$) to vary. 

% Pierre:
% The synchrotron model is calculated according to the method presented in Ginzburg (1979 ,"Theor. Phys. & Astrophys.") and Ghisellini, Guilbert, Svensson (1988, ApJ 334, L5 eq.[9]).
% The values 0.22 - 1.5 mG from Brogan et al. (2000) are upper limits in maser regions where the compression of the shocked gas increases the magnetic field - the synchrotron emission is diffuse
% A cut off energy < 1.5 TeV (exponential cutoff) should be taken into account else the synchrotron flux in the X-ray domain would be larger the X-ray flux measured by XMM.

As synchrotron emission can occur at wavelengths up to X-ray, we examined data from the XMM-Newton observatory. Using approximately $17$\,ks of XMM-Newton observation from 1 March 2006 (ObsID 0306510101), we analyzed the data with the standard XMM-Newton Science Analysis Software (SAS v.9.0)\footnote{Available at \url{http://xmm.esac.esa.int/xsa}.}. %\fnref{xmmSAS}. 
We extracted the spectral flux from the region of excess X-ray flux coincident with the Fermi-LAT source. 
H.E.S.S. analysis of the same data and of Chandra data in the same region found the spectra to be compatible with absorbed thermal emission \citep{HESS_CTBA}. Our XMM-Newton analysis yields a flux likewise consistent with an absorbed thermal spectrum with the same order parameters as those given for the Chandra data: a temperature of several hundred eV and column density of a few$\,\times 10^{22}\,\mathrm{cm}^{-2}$. We suspect we slightly underestimated the parameters relative to those found for Chandra as we extracted the spectrum under the point source assumption.%, leading to a decrease in overall flux
\fntext[xmmSAS]{Available at \url{http://xmm.esac.esa.int/xsa}.}

Assuming the X-ray emission is thermal, the (lowest energy) x-ray data thus constrains the highest energy synchrotron emission the lepton population may produce: for the reasonable model tested herein, a cutoff energy greater than $\sim 1.5\,$TeV would produce an excess of synchrotron emission not observed in X-rays. In a similar fashion, for this model and fixed set of parameters, we can constrain $\mathrm{E}_{\mathrm{e,cut}} \lesssim 50\,$GeV to avoid excessive flux from bremsstrahlung emission in the H.E.S.S. domain. We anticipate similar constraints from performing a full fit of all the parameters.

% We ensured that the modeled synchrotron emission did not exceed the thermal X-ray emission measured with XMM-Newton.

%***do the relative sizes really make this realistic?? (Rather than: We extracted the location and extension of the previously-identified extended emission \cite{HESS_CTBA} by fitting Gaussians the extended excess emission. To form the spectra, we used the $2\sigma$ or $95\%$ containment region and removed instrumental lines.)

%\section{Discussion}\label{sec:Disc}

\section{Conclusions}\label{sec:Conclusions}
We robustly detect a source coincident with SNR CTB 37A at $18.6\sigma$ using the Fermi-LAT. 
Fermi-LAT $\gamma$-ray data, in concert with radio, X-ray, and TeV data, allows us to determine the type of emission and its characteristics by determining the most likely model.  %, as well as to constrain local environment parameters
In this work, we are able to fit the multiwavelength emission with a reasonable model combining emission from hadronic and leptonic populations. 
Thus, SNR CTB 37A is a potential CR accelerator.

We are currently finishing a complete fit of the multiwavelength data over the allowable ranges for the local magnetic field and the leptonic and hadronic populations' parameters to determine their most probable values. In particular, this method permits us to robustly determine whether one population's emission dominates over the other or if both are necessary to reproduce the observed data. Since our initial model with reasonable values reproduced the Fermi-LAT data moderately well, we anticipate similar values for the more robust fit. 

Additional sources of data, such as microwave (e.g. Planck), infrared (e.g. Spitzer), and optical (which we will obtain when the source reemerges from behind bright solar system objects) %the sun -- and the moon! 
 data may further help disentangle the emission models. By assembling individual CR source candidates such as SNRs into statistically significant populations, we will improve our understanding of the potential source classes, allowing comparison to properties derived from direct CR detection experiments, and more fully illuminating a hundred-year mystery. 

\section{Acknowledgements}
The \textit{Fermi} LAT Collaboration acknowledges generous ongoing support
from a number of agencies and institutes that have supported both the
development and the operation of the LAT as well as scientific data analysis.
These include the National Aeronautics and Space Administration and the
Department of Energy in the United States, the Commissariat \`a l'Energie Atomique
and the Centre National de la Recherche Scientifique / Institut National de Physique
Nucl\'eaire et de Physique des Particules in France, the Agenzia Spaziale Italiana
and the Istituto Nazionale di Fisica Nucleare in Italy, the Ministry of Education,
Culture, Sports, Science and Technology (MEXT), High Energy Accelerator Research
Organization (KEK) and Japan Aerospace Exploration Agency (JAXA) in Japan, and
the K.~A.~Wallenberg Foundation, the Swedish Research Council and the
Swedish National Space Board in Sweden.

Additional support for science analysis during the operations phase is gratefully
acknowledged from the Istituto Nazionale di Astrofisica in Italy and the Centre National d'\'Etudes Spatiales in France.


\begin{thebibliography}{18}
\expandafter\ifx\csname natexlab\endcsname\relax\def\natexlab#1{#1}\fi
\expandafter\ifx\csname url\endcsname\relax
  \def\url#1{\texttt{#1}}\fi
\expandafter\ifx\csname urlprefix\endcsname\relax\def\urlprefix{URL }\fi

\bibitem[{{Abdo} et~al.(2010{\natexlab{a}}){Abdo}, {Ackermann}, {Ajello}}]{1FGLCat}
{Abdo}, A.~A., {Ackermann}, M., {Ajello}, M., et~al. Jun. 2010{\natexlab{a}}. {Fermi Large Area Telescope First Source Catalog}. \apjs  188, 405--436.

\bibitem[{{Abdo} et~al.(2010{\natexlab{b}}){Abdo}, {Ackermann}, {Ajello},
  {Atwood}, {Axelsson}, {Baldini}, {Ballet}, {Barbiellini}, {Baring},
  {Bastieri}, and et~al.}]{PulsarCatalog}
{Abdo}, A.~A., {Ackermann}, M., {Ajello}, M., et~al. Apr. 2010{\natexlab{b}}. {The First Fermi Large Area Telescope Catalog of Gamma-ray Pulsars}. \apjs 187, 460--494.

\bibitem[{{Abdo} et~al.(2010{\natexlab{c}}){Abdo}, {Ackermann}, {Ajello}}]{JurgenLMC}
{Abdo}, A.~A., {Ackermann}, M., {Ajello}, M., et~al.
  Mar. 2010{\natexlab{c}}. {Observations of the Large Magellanic Cloud with
  Fermi}. \aap 512, A7+.

\bibitem[{{Abdo} et~al.(2010{\natexlab{d}}){Abdo}, {Ackermann}, {Ajello},
  {Baldini}, {Ballet}, {Barbiellini}, {Bastieri}, {Baughman}, {Bechtol}}]{IC443LAT}
{Abdo}, A.~A., {Ackermann}, M., {Ajello}, M., et~al.
 Mar. 2010{\natexlab{d}}.  {Observation of Supernova Remnant IC 443 with the Fermi Large Area
  Telescope}. \apj 712, 459--468.

\bibitem[{{Aharonian} et~al.(2008){Aharonian}, {Akhperjanian}, {Barres de Almeida}, {Bazer-Bachi}}]{HESS_CTBA}
{Aharonian}, F., {Akhperjanian}, A.~G., {Barres de Almeida}, U., et~al. Nov. 2008.
  {Discovery of a VHE gamma-ray source coincident with the supernova remnant CTB 37A}. \aap 490, 685--693.

\bibitem[{{Atwood} et~al.(2009){Atwood}, {Abdo}, {Ackermann}, {Althouse}, {Anderson}}]{AtwoodLAT}
{Atwood}, W.~B., {Abdo}, A.~A., {Ackermann}, M., et~al. Jun. 2009. {The Large Area Telescope on the
  Fermi Gamma-Ray Space Telescope Mission}. \apj 697, 1071--1102.

\bibitem[{{Blumenthal} and {Gould}(1970)}]{BlumenthalGould}
{Blumenthal}, G.~R., {Gould}, R.~J., 1970. {Bremsstrahlung, Synchrotron
  Radiation, and Compton Scattering of High-Energy Electrons Traversing Dilute
  Gases}. Reviews of Modern Physics 42, 237--271.

\bibitem[{{Brandt}(2010)}]{myThesis}
{Brandt}, T.~J., 2010. {On high energy cosmic rays from the CREAM instrument}.
  Ph.D. thesis, The Ohio State University.

\bibitem[{{Brogan} et~al.(2000){Brogan}, {Frail}, {Goss}, and
  {Troland}}]{BroganZeeman}
{Brogan}, C.~L., {Frail}, D.~A., {Goss}, W.~M., {Troland}, T.~H., Jul. 2000.
  {OH Zeeman Magnetic Field Detections toward Five Supernova Remnants Using the
  VLA}. \apj 537, 875--890.

\bibitem[{{Castro} and {Slane}(2010)}]{CastroSlane2010}
{Castro}, D. and {Slane}, P., Jul. 2010. {Fermi Large Area Telescope Observations of Supernova Remnants Interacting with Molecular Clouds}. \apj 717, 372-378.

\bibitem[{{Dormody} et~al.(2011)}]{BlindPulseSearch}
{Dormody}, M., et~al., 2011. {In preparation}.

\bibitem[{{Frail} et~al.(1996){Frail}, {Goss}, {Reynoso}, {Giacani}, {Green},
  and {Otrupcek}}]{Frail96Masers}
{Frail}, D.~A., {Goss}, W.~M., {Reynoso}, E.~M., {Giacani}, E.~B., {Green},
  A.~J., {Otrupcek}, R., Apr. 1996. {A Survey for OH (1720 MHz) Maser Emission
  Toward Supernova Remnants}. \aj 111, 1651--+.

\bibitem[{{Ghisellini} et~al.(1988){Ghisellini}, {Guilbert}, and
  {Svensson}}]{GhiselliniGuilbertSvensson}
{Ghisellini}, G., {Guilbert}, P.~W., {Svensson}, R., Nov. 1988. {The
  synchrotron boiler}. \apjl 334, L5--L8.

\bibitem[{{Ginzburg}(1979)}]{Ginzburg}
{Ginzburg}, V.~L., 1979. {Theoretical physics and astrophysics}. International
  Series in Natural Philosophy, Oxford: Pergamon.

\bibitem[{{Green}(2009)}]{GreensCat}
{Green}, D.~A., Mar. 2009. {A revised Galactic supernova remnant catalogue}.
  Bulletin of the Astronomical Society of India 37, 45--+, available at
  \url{http://www.mrao.cam.ac.uk/surveys/snrs/}.

\bibitem[{{Kamae} et~al.(2006){Kamae}, {Karlsson}, {Mizuno}, {Abe}, and
  {Koi}}]{Kamae2006}
{Kamae}, T., {Karlsson}, N., {Mizuno}, T., {Abe}, T., {Koi}, T., Aug. 2006.
  {Parameterization of {$\gamma$}, $e^{+/-}$, and Neutrino Spectra Produced by
  p-p Interaction in Astronomical Environments}. \apj 647, 692--708.

\bibitem[{{Kassim} et~al.(1991){Kassim}, {Weiler}, and {Baum}}]{Kassim91}
{Kassim}, N.~E., {Weiler}, K.~W., {Baum}, S.~A., Jun. 1991. {A new look at the
  'jet' in the CTB 37A/B supernova remnant complex}. \apj 374, 212--217.

\bibitem[{{Porter} et~al.(2008){Porter}, {Moskalenko}, {Strong}, {Orlando}, and
  {Bouchet}}]{Porter2008}
{Porter}, T.~A., {Moskalenko}, I.~V., {Strong}, A.~W., {Orlando}, E.,
  {Bouchet}, L., Jul. 2008. {Inverse Compton Origin of the Hard X-Ray and Soft
  Gamma-Ray Emission from the Galactic Ridge}. \apj 682, 400--407.

\bibitem[{{Reynoso} and {Mangum}(2000)}]{ReynosoMagnumCO}
{Reynoso}, E.~M., {Mangum}, J.~G., Dec. 2000. {CO Observations toward Supernova
  Remnants with Associated OH 1720 MHZ Masers}. \apj 545, 874--884.

\end{thebibliography}
\end{document}